\documentclass[12pt,preprint]{aastex}

\usepackage{graphicx}

\shorttitle{FU Ori model} \shortauthors{Zhu et al.}

\begin{document}

\title{THE HOT INNER DISK OF FU ORI}

\author{Zhaohuan Zhu\altaffilmark{1}, Lee Hartmann\altaffilmark{1}, Nuria Calvet\altaffilmark{1},Jesus Hernandez\altaffilmark{1,3}, James
Muzerolle\altaffilmark{2}, Ajay-Kumar Tannirkulam\altaffilmark{1}}

\altaffiltext{1}{Dept. of Astronomy, University of Michigan, 500
Church Street, Ann Arbor, MI
48109;zhuzh@umich.edu,lhartm@umich.edu,ncalvet@umich.edu,hernandj@umich.edu,hernandj@umich.edu}
\altaffiltext{2}{Steward Observatory, 933 North Cherry Avenue,
University of Arizona,Tucson, AZ 85721;jamesm@as.arizona.edu}
\altaffiltext{3}{Centro de Investigaciones de Astronomia, Apartado
Postal 264, Merida 5101-A, Venezuela} \email{}

\newcommand\msun{\rm M_{\odot}}
\newcommand\lsun{\rm L_{\odot}}
\newcommand\rsun{\rm R_{\odot}}
\newcommand\msunyr{\rm M_{\odot}\,yr^{-1}}
\newcommand\be{\begin{equation}}
\newcommand\en{\end{equation}}
\newcommand\cm{\rm cm}
\newcommand\kms{\rm{\, km \, s^{-1}}}
\newcommand\K{\rm K}
\newcommand\etal{{\rm et al}.\ }
\newcommand\sd{\partial}

\begin{abstract}
We have constructed a detailed radiative transfer disk model which
reproduces the main features of the spectrum of the outbursting
young stellar object FU Orionis from $\sim$ 4000 \AA\, to $\sim 8
\mu$m. Using an estimated visual extinction $A_{V} \sim 1.5$, a
steady disk model with a central star mass $\sim 0.3 M_{\odot}$ and
a mass accretion rate $\sim 2 \times 10^{-4} M_{\odot}$yr$^{-1}$, we can
reproduce the spectral energy distribution of FU Ori quite well.
Higher values of extinction used in previous analysis (A$_{V}$$\sim$ 2.1) result
in spectral energy distributions which are less well-fitted by a
steady disk model, but might be explained by extra energy
dissipation of the boundary layer in the inner disk. With the
mid-infrared spectrum obtained by the Infrared Spectrograph (IRS) on
board the {\em Spitzer Space Telescope}, we estimate that the outer
radius of the hot, rapidly accreting inner disk is $\sim$ 1 AU using disk
models truncated at this outer radius. Inclusion of radiation from
a cooler irradiated outer disk might reduce the outer limit of
the hot inner disk to $\sim$ 0.5 AU.
In either case, the radius is inconsistent with a pure thermal instability model for the
outburst. Our radiative transfer model implies that the central disk
temperature $T_{c} \ge 1000$~K out to $\sim$ 0.5 $-$ 1 AU, suggesting that
the magnetorotational instability can be supported out to that
distance. Assuming that the $\sim 100$~yr decay timescale in
brightness of FU Ori represents the viscous timescale of the hot
inner disk, we estimate the viscosity parameter to be $\alpha \sim
0.2 - 0.02$ in the outburst state, consistent with numerical
simulations of the magnetorotational instability in disks. The
radial extent of the high $\dot{M}$ region is inconsistent with the
model of Bell \& Lin, but may be consistent with theories
incorporating both gravitational and magnetorotational
instabilities.
\end{abstract}

\keywords{accretion disks, circumstellar matter, stars: formation, stars: pre-main sequence}

\section{Introduction}
The FU Orionis systems are a small but remarkable class of variable young objects
which undergo outbursts in optical light of 5 magnitudes or more \citep{herbig77}, with a F-G
supergiant optical spectra and K-M supergiant
near-infrared spectra dominated by deep CO overtone absorption.
While the rise times for outbursts are usually very short ($\sim$ 1-10 yr), the
decay timescales range from decades to centuries.  The FU Ori
objects also show distinctive reflection nebulae, large infrared
excesses of radiation, wavelength dependent spectral types, and ``double-peaked''
absorption line profiles \citep{Lee96}. The frequency of these
outbursts is uncertain; in recent years an increasing number of heavily extincted
potential FU Ori objects have been identified on the basis of their spectroscopic
characteristics at near-infrared wavelengths \citep{kenyon93,reipurth97,sandell98,aspin03}.

The accretion disk model for FU Ori objects proposed by Hartmann
$\&$ Kenyon (1985, 1987a, 1987b) and \cite{kenyon88} can explain the
peculiarities enumerated above in a straightforward manner.
Outbursts are known in other accreting disk systems and may be the
result of a common mechanism (e.g., Bell $\&$ Lin 1994). The high
temperature of the inner disk produces the observed F-G supergiant
optical spectrum, while the cooler outer disk produces an infrared
spectrum having the spectral type of a K-M supergiant. The Keplerian
rotation of the disk can produce double-peaked line profiles as
often observed, with peak separation decreasing with increasing
wavelength of observations, since the inner hotter disk which
produces the optical spectrum rotates faster than the outer cooler
disk which produces the infrared spectrum \citep{Lee96}.

While the accretion disk model has been successful so far, it is
important to continue to test it and to derive further insights into
the origins of the accretion outbursts as new observations become
available.  In particular, the mid-infrared spectrum obtained with
the Infrared Spectrograph (IRS) on the {\em Spitzer Space Telescope}
provides important constraints on the outer edge of the hot inner
disk \citep{jgreen06}, which in turn can test theories of outbursts.
For example, the thermal instability model of \cite{bell94} predicts
an outer radius for the outburst region of only $\sim 20 \rsun$,
which is smaller than that estimated by \cite{jgreen06} on the basis
of simple blackbody disk models. \cite{armitage01} suggested that a
combination of gravitational instability (GI) with triggering of the
magnetorotational instability (MRI) might also explain FU Ori
outbursts; in this model the high accretion rate region in outburst
would be much larger, of order 0.5 AU. Vorobyov \& Basu (2005,2006)
suggested that FU Ori outbursts could also be produced by the
accretion of clumps formed in a gravitationally unstable disk. This
model predicts high accretion rates at its inner radius of 10 AU.
These differing predictions for the size of the outburst region
should be testable with detailed models of the spectral energy
distribution (SED).
 Furthermore, if we can
constrain the size of the outburst region, the timescales of decay
may provide quantitative estimates of the viscosity, vital for
understanding the evolution of young protoplanetary disks.

In this paper we present new steady, optically-thick accretion disk spectrum calculations,
and compare the results with observations from the ultraviolet
to the mid-infrared of FU Ori$-$the prototype of these systems.
In \S 2 we describe the observational material used to constrain the models, while
in \S 3 we describe the methods used to calculate the disk models.
We show in \S 4 that the steady disk model reproduces the variation of the observed
spectral features over this very large wavelength range quite well.
In \S 5 we consider some implications of our results for disk viscosities, outburst
mechanisms, and disk masses, and summarize our conclusions in \S 6.

\section{Optical and infrared data}

An essential part of the spectral energy distribution (SED) of FU Ori for our model is the
 IRS spectrum from {\em Spitzer} observed on March 4, 2004 (see Green et al. 2006 for details).
As FU Ori has been fading slowly over the last 60 years, we need to obtain data at other
wavelengths near the same time as the IRS spectrum to construct a complete spectral
energy distribution (SED).

Optical photometry
(UBVR) was obtained in 2004 at the Maidanak Observatory
\citep{ibrahimov99,jgreen06}.  As FU Ori exhibits small, irregular variability
on timescales much less than a year, we averaged the Maidanak data
to form the mean values (with RMS uncertainty in each band $\sim 0.02$ magnitude)
used in this paper.  In addition, we include an optical spectrum of
FU Ori obtained on March 27th, 2004, near
the time of the IRS observations, using the 1.5m telescope of the
Whipple Observatory with the FAST Spectrograph \citep{fabricant98}, equipped
with the Loral 512 $\times$ 2688 CCD. This instrument provides
3400 {\AA} of spectral coverage centered
at 5500 {\AA}, with a resolution of 6 \AA. The spectrum was wavelength calibrated
and combined using standard IRAF
routines \footnote{IRAF is distributed by the National Optical Astronomy Observatories,
which are operated by the Association of Universities for Research in Astronomy, Inc.,
under cooperative agreement with the National Science Foundation.}. The spectrum was
corrected for the relative system response using the IRAF  {\it
sensfunc} task.  The optical spectrum was placed on an absolute flux scale
using the averaged 2004 UBVR photometry.

Near-infrared fluxes were determined from several sources. 2MASS JHK
photometry is available from October 7th, 1999, five years before
IRS observations.  We also include data from \cite{reipurth2004} on
December 15, 2003, using the NAOJ 8 m Subaru Telescope on Mauna Kea
and the Infrared Camera and Spectrograph (IRCS). Finally, we
obtained JHK magnitudes on December 17th 2005, using the Near-IR
Imager/Spectrometer TIFKAM on the 2.4 m MDM telescope at Kitt Peak.
Although the 2.4m telescope had to be defocused
to avoid saturation because FU Ori is so bright,
our JHK magnitudes of 6.57, 5.91 and 5.33, respectively,
are in reasonable agreement with the JHK$_s$ 2MASS magnitudes of
6.519, 5.699, 5.159, respectively, indicating that the FU Ori
fluxes have not changed dramatically over that six year period.
The \cite{reipurth2004} observations are on a different system,
but when converted to fluxes yield reasonably similar results.
We therefore adopted the 2MASS magnitudes as our standard
in this wavelength region.

We then incorporated the KSPEC spectrum (resolution R$\sim$500,
1.15-2.42$\mu$m ) observed on Dec 15th, 1994 by \cite{greene96} and
the observation with SpeX (2.1-4.8$\mu$m) on the Infrared Telescope
Facility (IRTF; Rayner et al. 1998) obtained during the nights of January
5-6th, 2001 (see Muzerolle et al. 2003 for details). We placed these
spectra on an absolute scale by convolving the spectra with the
relative spectral response curve (RSRs) of the 2MASS
system \citep{cohen03} and tying the overall result to the 2MASS
K$_s$ magnitude.

\begin{figure}
\epsscale{.80} \plotone{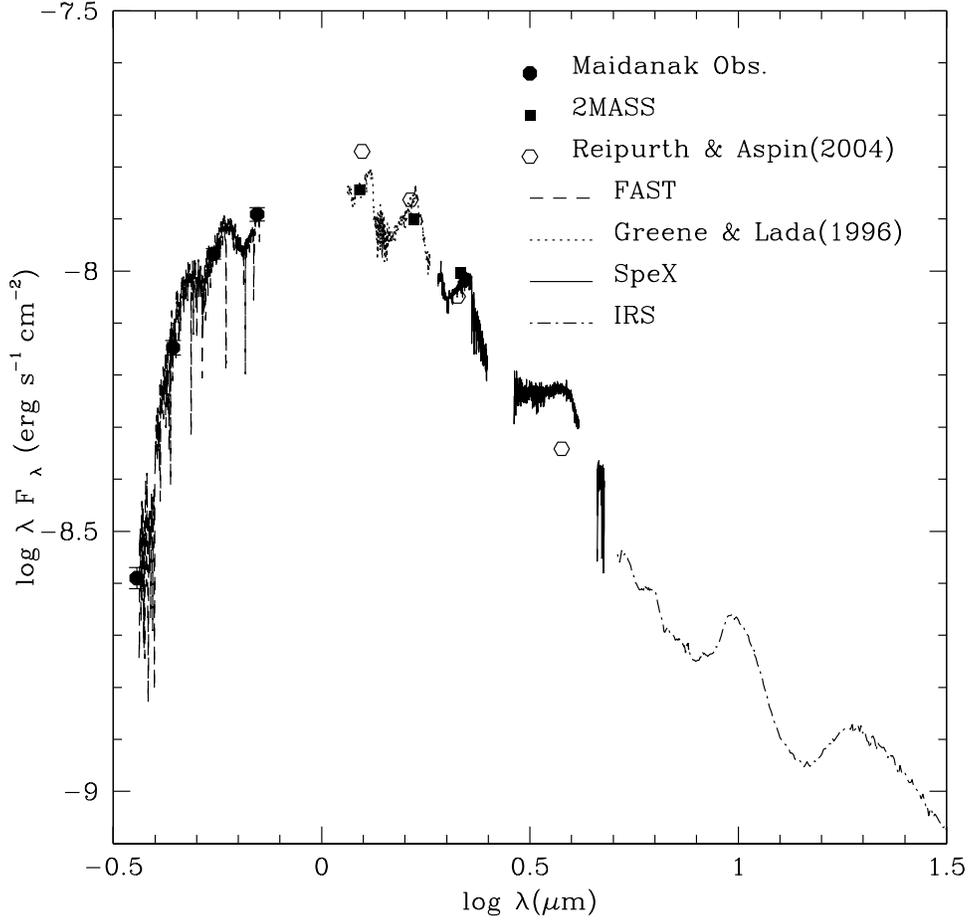} \caption{The spectral energy distribution of FU Ori.
Collected photometric and spectroscopic data of FU Ori with A$_{v}$=1.5 (see \S 4).
The solid circles with error bars are the UBVR photometry from the Maidanak Observatory,
the squares are 2MASS JHK photometry, and the open circles are JHK'L' photometry from \cite{reipurth2004}.
The dashed curve is the FAST spectrum from 3650 \AA \ to 7500 \AA,
the dotted curve is the KSPEC spectrum from 1.15 $\mu$m to 2.42 $\mu$m \citep{greene96},
and the solid curve is the SpeX spectrum from 2.1 $\mu$m to 4.8 $\mu$m
\citep{muzerolle03}, both scaled to the 2MASS photometry.
Finally, the dot-dash curve is the IRS spectrum from 5 $\mu$m to 35 $\mu$m
from \cite{jgreen06}. The spectra were put on as absolute flux scale as discussed in
\S 2. }
\label{fig:1}
\end{figure}

The collected observations are shown in Figure \ref{fig:1}, where we assume an A$_{V}$
of 1.5 instead of the usually assumed value of $\sim$ 2.2 (Kenyon \etal 1988; KHH88).
The previously discussed variation of spectral type with wavelength is immediately
apparent in Figure \ref{fig:1};
the optical spectrum corresponds to a G type spectrum as estimated
from the atomic metal lines while
the near-IR and IRS wavelength ranges are dominated by molecular bands indicative
of a much later spectral type.  While the optical and infrared spectra are
dominated by absorption features, dust emission dominates longward of
8 $\mu$m, as shown by the silicate emission features at 10 and 20 $\mu$m.

\section{Model}

In this paper we only model the absorption spectrum ($\lambda$$<$8 $\mu$m)
which is produced by the centrally-heated accretion disk.  The silicate emission
features are produced by externally-heated regions of the outer disk or possibly
a circumstellar dusty envelope (Kenyon \& Hartmann 1991; Green \etal 2006);
we defer modeling of this region to a subsequent paper.

We follow the method of Calvet \etal (1991a,b) to calculate the disk
spectrum. In summary, we calculate the emission from the atmosphere of
 a viscous, geometrically thin,
optically thick accretion disk with constant mass accretion rate
$\dot{M}$ around a star with mass M and radius R. The disk height
$H$ is assumed to vary with the distance from the rotational axis of
the star $R$ as $H=H_{0}(R/R_{i})^{9/8}$, where we take $H_{0}=0.1
R_{i}$ and $R_{i}$ is the radius of the central star. This
approximation is not very accurate but it only affects the local
surface gravity of the disk atmosphere, which has only a small
effect on the emergent spectrum. Viscous dissipation in the
atmosphere is neglected, which is a good approximation for cases of
interest \citep{HK91}. Thus, radiative equilibrium holds in the disk
atmosphere, and the surface flux is determined by the viscous energy
generation in the deeper disk layers.  This constant radiative flux
through the disk atmosphere can be characterized by the effective
temperature distribution of the steady optically-thick disk,
\begin{equation}
T_{eff}^{4}= \frac{3GM\dot{M}}{8\pi\sigma R^{3}} \left [ 1-\left (
\frac{R_{i}}{R} \right ) ^{1/2} \right ] \,. \label{eq:teff}
\end{equation}
This equation predicts that the maximum disk temperature $T_{max}$
occurs at $1.36 R_{i}$ and then decreases to zero at $R = R_i$.  The
latter is unphysical, and so we modify equation (\ref{eq:teff}) so
that when the radius is smaller than 1.36 R$_{i}$, we assume that
the temperature is constant and equal to $T = T_{max}$. (e.g.,
 KHH88) The vertical temperature structure at each
radius is calculated using the gray-atmosphere approximation in the
Eddington limit, adopting the Rosseland mean optical depth $\tau$.

The opacity of atomic and molecular lines has been calculated using
the Opacity Distribution Function (ODF) method \citep{kurucz04,
kurucz042,castelli05}. Briefly, the ODF method is a statistical
approach to handling line blanketing when millions of lines are
present in a short wavelength range \citep{kurucz74}. For a
given temperature and pressure, the absorption coefficient for each
line is exactly computed, then the profiles of all the lines in each
small interval $\Delta \nu$ are rearranged monotonically as a
function of $\nu$. The opacity increases as the frequency increases
in this $\Delta \nu$. A step function with 12 subintervals in
frequency is used to represent this monotonic function. The height
of every step is the averaged opacity of the monotonic function in
this subinterval. The width of the 12 subintervals are $\Delta
\nu$/10 for 9 intervals and $\Delta \nu$/20, $\Delta \nu$/30,
$\Delta \nu$/60 for the last 3 intervals where the monotonic
function increases steeply and reaches the maximum.  Thus, in every
interval $\Delta \nu$ we obtain 12 representative averaged opacities
at a given temperature and pressure. We then construct an opacity
table as a function of $\lambda$ for each temperature and pressure.
The wavelength grid is the same as used in the code DFSYNTHE
\citep{castelli05} from 8.976 nm to 10000 nm with 328 BIG intervals
(resolution ~30-100) and 1212 LITTLE intervals (resolution ~60-500).
However, we extend the temperature grid and pressure grid to lower
temperatures and pressures ($\log T$ ranging from 1.5 to 5.3, $\log
P$ from -24 to 8 in cgs units) than DFSYNTHE, so that it can be used in our disk
models for which the temperatures and pressures reach lower values
than in typical stellar atmosphere models. Because the line opacity
usually varies by several orders of magnitude within a line width,
the ODF method is substantially more accurate than either straight
or harmonic means.

The line list is taken from Kurucz's CD-ROMs Nos. 1, 15, 24, and 26
\citep{kurucz05}.  Not only atomic lines but also many molecular
lines (C$_{2}$, CN, CO, H$_{2}$, CH, NH, OH, MgH, SiH, SiO, TiO,
H$_{2}$O) are included. The opacities of TiO and H$_{2}$O, the most
important molecules in the infrared, are from \cite{PS97} and
\cite{S98}. We do not include a detailed calculation of the dust
condensation process; instead, we add the dust opacity when T$<$1500
K (the condensation temperature of silicates at typical disk
densities). We use the ISM dust model of \cite{draine84,draine87} to
represent the dust in the disk. This assumption is unlikely to be
correct but mostly affects the continuum at the very longest
wavelengths and in particular the silicate emission features, which
 we do not model in this paper.

At low temperatures complex chemical processes occur which are not included in the Kurucz data.
We have not calculated the low temperature
molecular opacity in detail;
instead, we have assumed that the abundance ratio between different types
of molecules below 700 K is the same as the ratio at 700 K. This is unimportant
for our purposes because dust opacity dominates at such low temperatures.

\begin{figure}
\epsscale{.80} \plotone{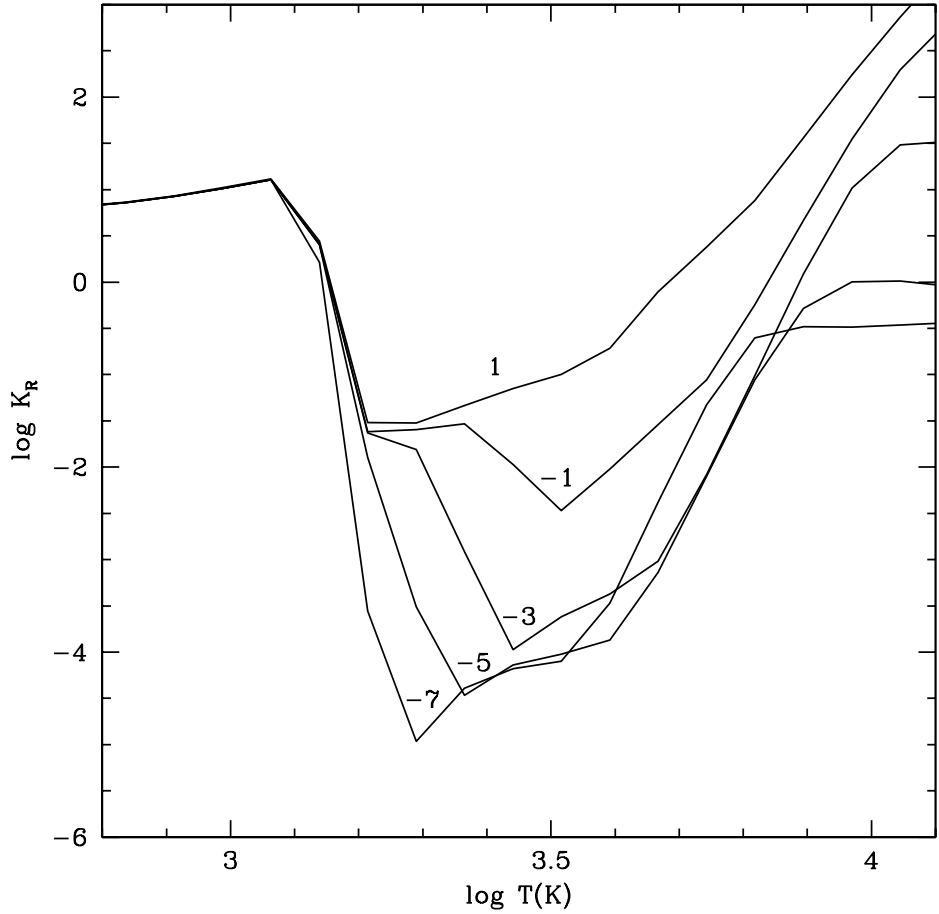} \caption{Our calculated Rosseland mean opacity
as a function of temperature for solar composition. Each curve is
labeled with the value of log($\rho /T_6^3$) to be compared
 with \cite{alexander94}, where $T_6$ is the temperature in millions of degrees. }
\label{fig:ross}
\end{figure}

In Figure \ref{fig:ross} we show the Rosseland mean opacity
calculated for solar abundances and a turbulent velocity $v = 2
\kms$ for different values of $\rho /T_6^3$ to be compared with
\cite{alexander94}, where $\rho$ is the density in gcm$^{-3}$ and
$T_6$ is the temperature in millions of degrees. The results closely
match the more detailed calculations by \cite{alexander94} and
\cite{ferguson05} except near 1500 K, where our dust opacity rises
more rapidly due to our neglect of dust condensation processes.

\begin{figure}
\epsscale{.80} \plotone{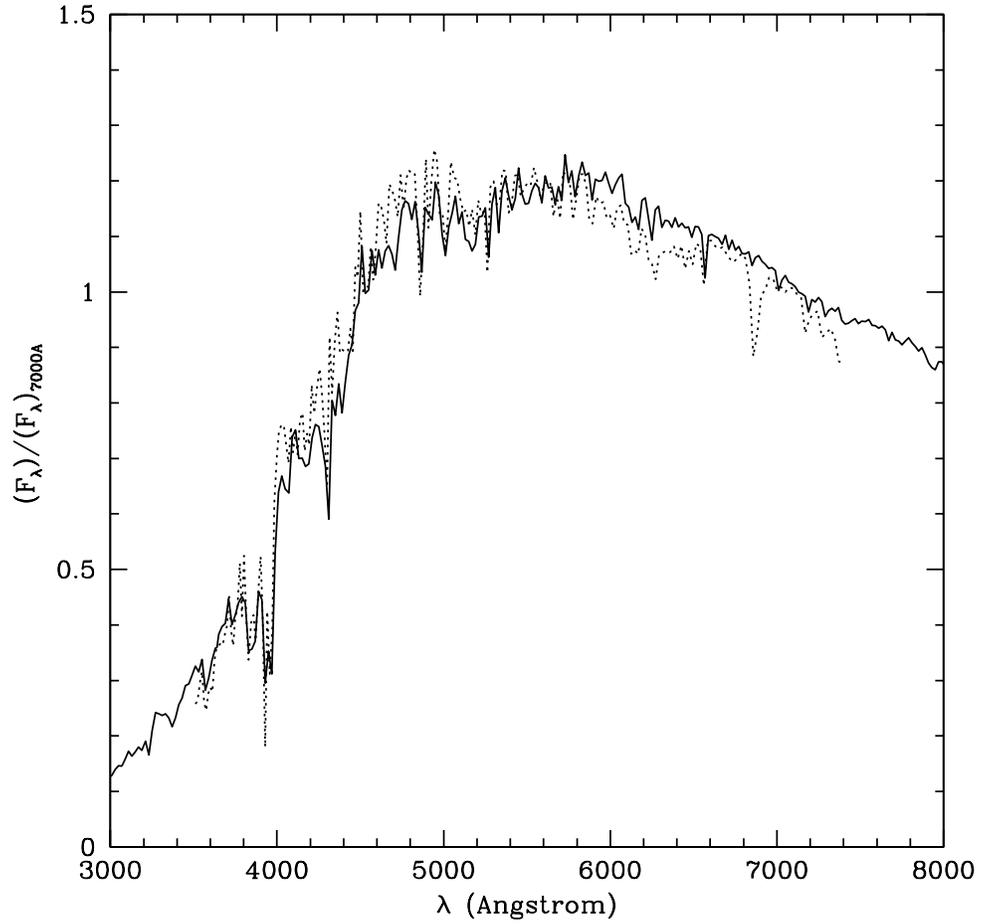} \caption{The spectrum predicted by
the annulus with T$_{eff}$=5300 K (solid line) and the spectrum of
SAO 21446 which is a G1 I star \citep{jacoby84} ( dotted line).
The spectra are scaled to the same flux at 7000 \AA. }
\label{fig:f3}
\end{figure}

\cite{calvet91b} showed that the near-infrared spectrum of FU Ori
could be well-modeled with the then-current water vapor opacities
and a similar calculational method to the one we are using here.
We now have a much better set of opacities, and can treat the optical spectrum.
 To test the code at optical wavelengths, we
compare the model spectrum from the annulus with T$_{eff}$=5300 K
with the observed spectrum of SAO 21446 which is a G1 supergiant
\citep{jacoby84}. The spectrum of SAO 21446 has been obtained using
the Intensified Reticon Scanner(IRS) on the No.1 90 cm telescope at
Kitt Peak by \cite{jacoby84}. Here we have convolved this spectrum
to a resolution of 300 to agree with our model spectrum.
As shown in Figure \ref{fig:f3},
our model reproduces the continuum spectrum fairly well
except for a few strong features. These discrepancies are probably
 due to the limitation of the grey atmospheric assumption.

\begin{figure}
\epsscale{.80}
\plotone{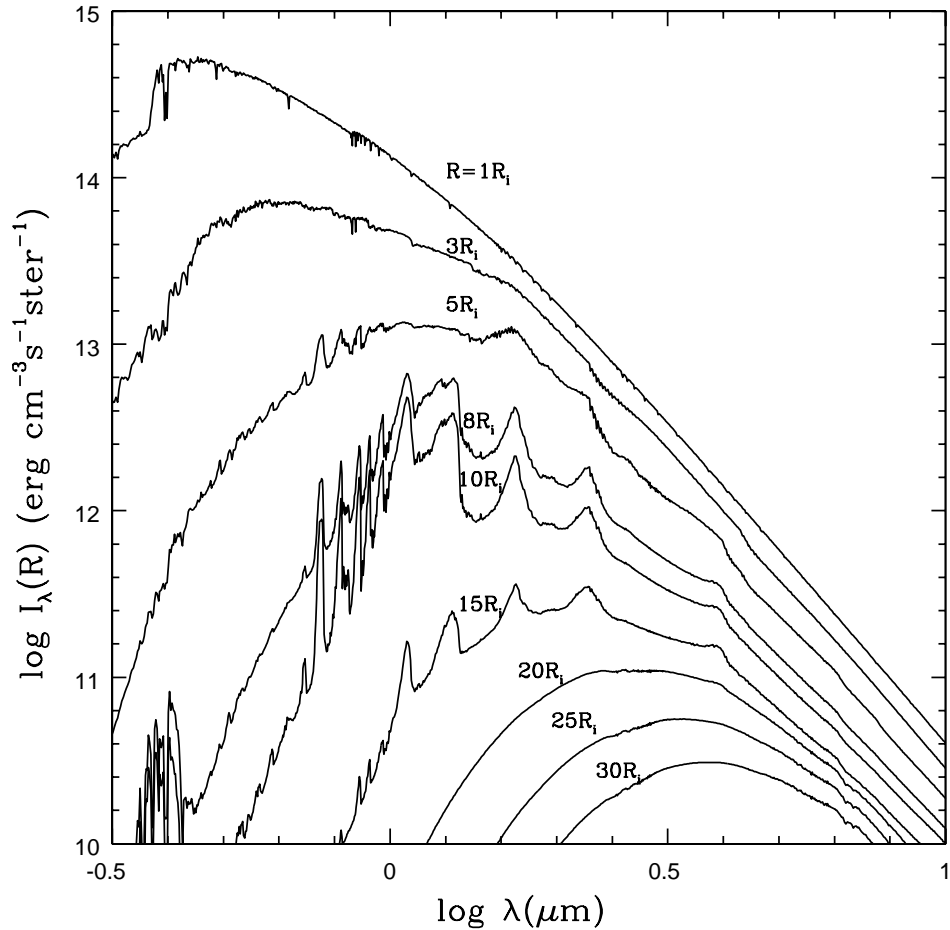}
\caption{Emergent intensities from selected annuli of the disk model calculated for a 55 degree emergent angle.
Effective temperatures of the annuli at 1, 3, 5, 6, 8, 10, 15, 20, 25, and 30 R$_{i}$
are 6420, 4660, 3400, 2480, 2130, 1630, 1310, 1110, and 977 K, respectively.
Molecular absorption features in the near-IR arise from cool regions of the disk,
which do not contribute to the optical spectrum.}
\label{fig:f4}
\end{figure}

Figure \ref{fig:f4} shows the emergent intensities with 55$^{o}$ emergent angle
from different annuli of
our final disk model (parameters can be found in \S 4), which has
45 annuli. The 55$^{o}$ inclination angle is estimated from the near-infrared interferometric
observations by \cite{malbet05} and mid-infrared interferometric observations by Quanz \etal (2006).
 The radii of these annuli are chosen to
increase exponentially from $r = R_i$ to $1000 R_i$.
At the first annulus ($r =1 R_i$), we can see a significant Balmer jump at 0.36 $\mu$m, similar to
the spectra of early type stars.  As the radius increases, the temperatures of the annuli
become lower and lower. Infrared molecular features appear at larger annuli with
effective temperatures less than 5000K ($r > 3 R_i$). For the outermost annuli with effective
temperature less than 1500 K ($r > 15 R_i$), the molecular features have almost disappeared
because dust opacity dominates.
The final spectrum is the result of the addition of the fluxes from each of these annuli
weighted by the appropriate annular surface area.

In addition to these calculations aimed at comparing with low-resolution spectra,
we also calculated high-resolution spectra in restricted wavelength regions. For these calculations,
the vertical structure of every annulus is calculated as above,
and then this structure is imported into the program
SYNTHE \citep{kurucz81,kurucz93}. SYNTHE is a suite of programs which solve the
radiative transfer equations in LTE with a very high spectrum resolution ($\sim$500,000).
Each annular spectrum is broadened with the rotational profile appropriate to a Keplerian
disk and then the summed spectrum is calculated (e.g., KHH88).
These calculations are very time-consuming, and were used only
to produce a high resolution spectrum of a small wavelength range around 6170 \AA\
for the purpose of estimating the central mass, and for examining the CO first-overtone lines.

\section{Comparison with observations}

The SED of a steady, optically-thick disk model is determined by
two parameters: the product of the mass accretion
rate and the mass of the central star, $\dot{M} M$, and the inner radius $R_i$.
One observational constraint comes from the observed spectral lines and/or the peak
of the SED (after reddening correction), which determines the maximum temperature
of the steady disk model,
\begin{equation}
T_{max}=0.488 \left ( \frac{3GM\dot{M}}{8\pi R_{i}^{3}\sigma} \right
)^{1/4}\,.\label{eq:tmax}
\end{equation}
The other observational constraint is given by the true luminosity
of a flat disk L$_{d}$,
\begin{equation}
L_{d} = 2 \pi d^{2} {F \over \cos i} = { G M \dot{M} \over 2
R_i}\,,\label{eq:Ld}
\end{equation}
where $d$ is the distance of FU Ori,
$i$ is the inclination angle of the disk to the line of sight, and
$F$ is the observed total flux corrected for extinction.
If the distance and inclination are known, we can solve for the inner disk radius
and thus for the product $M \dot{M}$.  If we then further use the observed
rotational velocities, it is then possible to derive independent values of the
central mass $M$ and the accretion rate $\dot{M}$, in the manner outlined by
KHH88.

\begin{figure}
\epsscale{.80}
\plotone{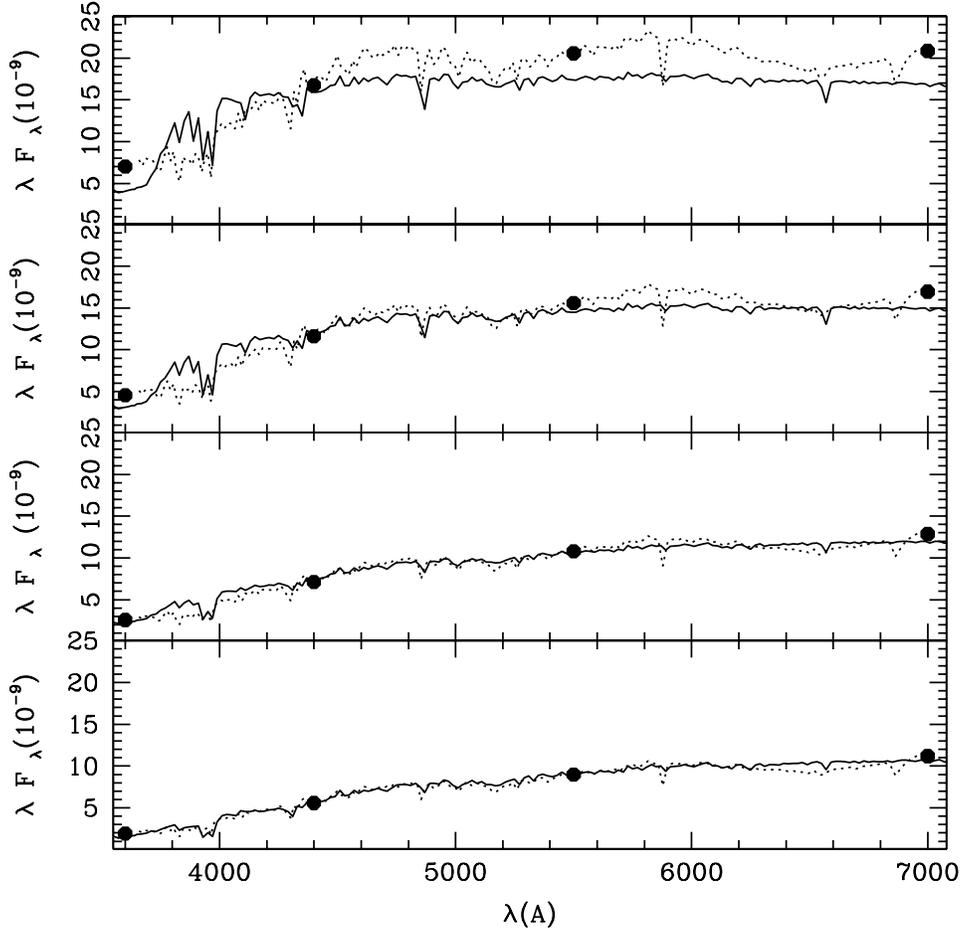}
\caption{Observed UBVR photometry and optical spectra
with different A$_{V}$: 2.2,1.9,1.5,1.3 (dotted lines, from top to bottom)
and model spectra with T$_{max}$: 7240, 6770, 6420, 5840
(solid lines, from top to bottom). The total flux spectra are produced by
weighting the intensities (Figure 4) by the appropriate annular surface areas.
} \label{fig:f5}
\end{figure}

The best constraints on $T_{max}$ come from the optical spectrum, as the
uncertain extinction makes it difficult to constrain by overall SED
fitting.  From the lines of our optical FAST spectrum
we derive a spectral type of $\sim$ G2 for FU Ori using the methods of \cite{jesus04}, which
agrees with previous results \citep{herbig77}.
However, it is not straightforward to derive a value of extinction and thus $T_{max}$
from this spectral type determination; the disk spectrum is not that of a single (standard)
star with a well-defined effective temperature, but instead it is a combination of hotter
and cooler regions, with an overall spectrum that clearly varies with the wavelength
of observation in a complicated way.
To address this problem, we constructed several disk models with a modest range of
$T_{max}$ and compared them with the observed optical spectrum dereddened by
differing amounts of visual extinction $A_V$, as shown in Figure \ref{fig:f5}.

The uppermost model in Figure \ref{fig:f5} has $T_{max} = 7240$~K,
comparable to that of the FU Ori disk model of KHH88.  The model
optical spectrum is too flat longward of $\sim 3900$~\AA\ in
comparison with the observations dereddened by the $A_V = 2.2$,
suggested by KHH88, or for $A_V=1.9$.  At the other extreme, we find
that the bottom model, with $T_{max} = 5840$~K, matches the observed
spectrum dereddened by only $A_V = 1.3$ quite well.  However, we
suspect that our gray-atmosphere approximation somewhat
underestimates the amount of line blanketing in the blue optical
region, especially shortward of the Ca II resonance lines at 3933
and 3968 \AA\, leading us to suspect that this agreement is somewhat
misleading. We therefore provide our best estimate of the extinction
as $A_V = 1.5 \pm 0.2$, and adopt the model which best matches the
mean extinction with $T_{max} = 6420$~K as our standard model.

In this connection it is worth noting that KHH88 did not actually
find a good fit for the optical spectrum of FU Ori using a GOI
standard star and extinctions $A_V > 2$.  They speculated that the
discrepancy might be alleviated by an improved treatment of limb
darkening in the disk.  However, Hartmann \& Kenyon (1985) had found
that a G2I standard star and $A_V = 1.55$ provided a good match to
the optical spectrum between 3900 \AA\ and 7400 \AA\,, consistent
with our results here.


The accretion disk luminosity $L$ depends upon the distance,
inclination, and dereddened flux.  We adopt a distance of $\sim$ 500
pc, consistent with membership in the general Orion region (Herbig
1977). The inclination 55$^{o}$ is adopted
(Malbet et al.2005, Quanz et al. 2006). Using these
parameters, observed total flux and $A_V = 1.5$, we obtain the true
luminosity $L_{d}=8.66\times10^{35} \rm ergs \,
s^{-1}$$\sim$226$L_{\odot}$ according to equation \ref{eq:Ld}. As
mentioned before, fitting the observed spectrum yields $T_{max} =
6420$~K. We derive $M\dot{M} = 7.2 \times 10^{-5}
M_{\odot}^{2}$yr$^{-1}$ and $R_i = 5 R_{\odot}$ by solving equation \ref{eq:tmax} and equation
\ref{eq:Ld} simultaneously. 

\begin{figure}
\epsscale{.80} \plotone{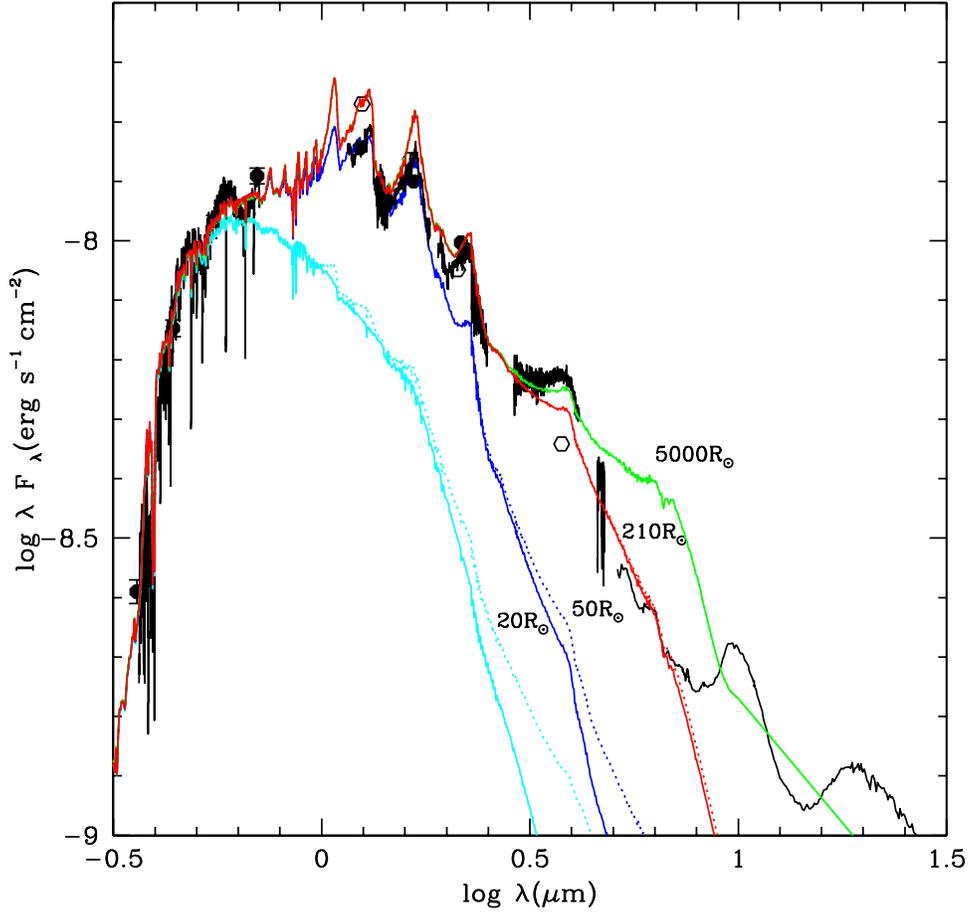} \caption{Comparison between
observed data and disk models for different values of the outer
radii. All the black lines and dots are observed spectra and
photometry data as shown in figure \ref{fig:1}. All the color lines
are model spectra. The parameters of the disk models are
 inner disk radius 5 R$_{\odot}$, and $M \dot{M}$$\sim$
7.2$\times$10$^{-5}$M$_{\odot}$/yr. The corresponding T$_{max}$ is
6420 K.
 The four model
spectra from top to bottom are for the disk models truncated at 5000
R$_{\odot}$, 210 R$_{\odot}$, 50 R$_{\odot}$ and 20 R$_{\odot}$. The
solid lines are model spectra for a high mass accretion rate region
with outer radii as shown. The dotted lines are spectra for models
where an outer region with mass accretion one order of magnitude
smaller has been added. For the disk model truncated at 5000
R$_{\odot}$, the dotted line is coincident with the solid line.
 The extinction parameter is $A_{V}$=1.5.
} \label{fig:f6}
\end{figure}

Figure \ref{fig:f6} shows the predicted spectrum for the adopted
parameters and several values of the outer radius.  Our computed
spectrum matches the observations at wavelengths
shortward of about $4 \mu$m, indicating that the innermost
region of the inner disk is reasonably well reproduced by the adopted
$\dot{M} M$ and $R_i$ parameters.

Because the optical and near-infrared SED is well-matched by the model, we may
then proceed to estimate the central stellar mass and thus the accretion rate.
We used the observed optical rotation from KHH88 to constrain the central star mass.
We computed the line profiles around 6170 $\dot{A}$ using SYNTHE, cross
correlated it with a non-rotating disk spectrum, and compared it with the
observed cross-correlation line profile given in KHH88. The uncertainty in
 fitting the line profile widths is about $\pm$~20\%. However, systematic
errors in the inclination angle are probably more important for the mass 
estimate, and for that reason we give an uncertainty 
 $0.3 M_{\odot}\pm0.1 M_{\odot}$ (
consistent with the earlier estimate of KHH88 assuming $i = 50^{\circ}$).
For a mass of 0.3 $M_{\odot}$, $\dot{M} \sim 2.4 \times 10^{-4} M_{\odot}$yr$^{-1}$.
Parameters of the model are given in Table 1.

As shown in Figure \ref{fig:f6},
we find that regardless of extinction and disk parameters, steady accretion models
which fit the optical to near-infrared region predict too much
emission in the IRS range for very large outer radii.  This can only be
remedied by truncating the hot inner disk.  The estimate of the
outburst models of \cite{bell94} suggested an outer radius of the
high state of $\sim 20 \rsun$ (with somewhat different parameters
for the accretion rate and inner disk radius); as shown in Figure
\ref{fig:f6} this truncation radius fails to explain the SED. A
truncation radius of $R_{out} \sim 210 R_{\odot} \sim 1$~AU provides
a better fit to the flux at $\sim 5 - 8 \mu$m, although is somewhat
low in the $3 \mu$m band.
Our results are reasonably consistent with
 \cite{jgreen06}, considering the limitations of their blackbody modeling only out to 5 microns.

As discussed above,
models with outer radii larger than $\sim$200$R_{\odot}$ overpredict the flux at $\sim 5 - 8 \mu$m.
At even longer wavelengths, absorption features are no longer present; instead,
silicate dust emission features are seen in IRS spectra at $\sim 10$ and $20 \mu$m.
As discussed by \cite{jgreen06}, these silicate features are signatures of heating
from above, rather than from internal viscous dissipation, and may be produced in
upper atmosphere of the outer disk which absorb light from the central disk.
An outer disk is expected to be present, since
a disk radius of $\sim 0.5 - 1$~AU  is extremely small by standards of typical T Tauri
disks, for which $R_{out} \sim 100$~AU or more are common (e.g., Simon, Dutrey,
\& Guilloteau 2000). Moreover,
such a small radius would imply a low-mass optically-thick disk,
making it difficult
to explain the observed submm flux from FU Ori
(Sandell \& Weintraub 2001). So, it is very likely that an outer disk is present with a lower
accretion rate, which could contribute some flux in the $5 - 8 \mu$m wavelength
range and thus reduce the radius of the hot, high-accretion rate region.
An outer disk accretion rate of about $10^{-5} \msunyr$ is comparable to some of
the highest infall rates estimated for (not heavily embedded) protostars (e.g., Kenyon,
Calvet \& Hartmann 1993); higher accretion rates would imply implausibly short
evolutionary timescales $M/\dot{M} \sim 3 \times 10^4$~yr.  In any event, if the
outer disk accretion rate is not significantly smaller than the inner disk accretion
rate, the problem of a large outer radius yielding too much flux remains.
More likely, the emission from the outer disk is dominated by irradiation from the
inner disk independent of the (lower) accretion rate there \citep{turner97}, as indicated
by the presence of emission rather than absorption features.

In the \cite{bell94} thermal instability models for FU Ori outbursts,
the outer disk accretion rate is of order $10^{-1}$ of the inner disk accretion
rate in outburst.  We have therefore investigated the effect of an outer disk
accretion rate of $2.4 \times 10^{-5} \msunyr$ on the SED.  The dotted curves
in Figure \ref{fig:f6} show the results of adding in such outer disk emission
 from disk internal viscous dissipation.
The effect is relatively small so that our estimate of the outer radius of
the hot disk with the high accretion rate is relatively robust.  Of course
we cannot rule out other kinds of temperature distributions, such as a smooth
decline of accretion rate from, say, 100 to 200 $\rsun$ and
the irradiation effect from the inner disk to the outer disk.
We are presently carrying out more detailed calculations of the outer
disk, including irradiation by the inner disk. Preliminary results indicate
that the addition of the irradiated outer disk emission
will decrease our hot disk outer radius estimate by no more than a factor of two (Zhu et al., in preparation).

\begin{figure}
\epsscale{.80} \plotone{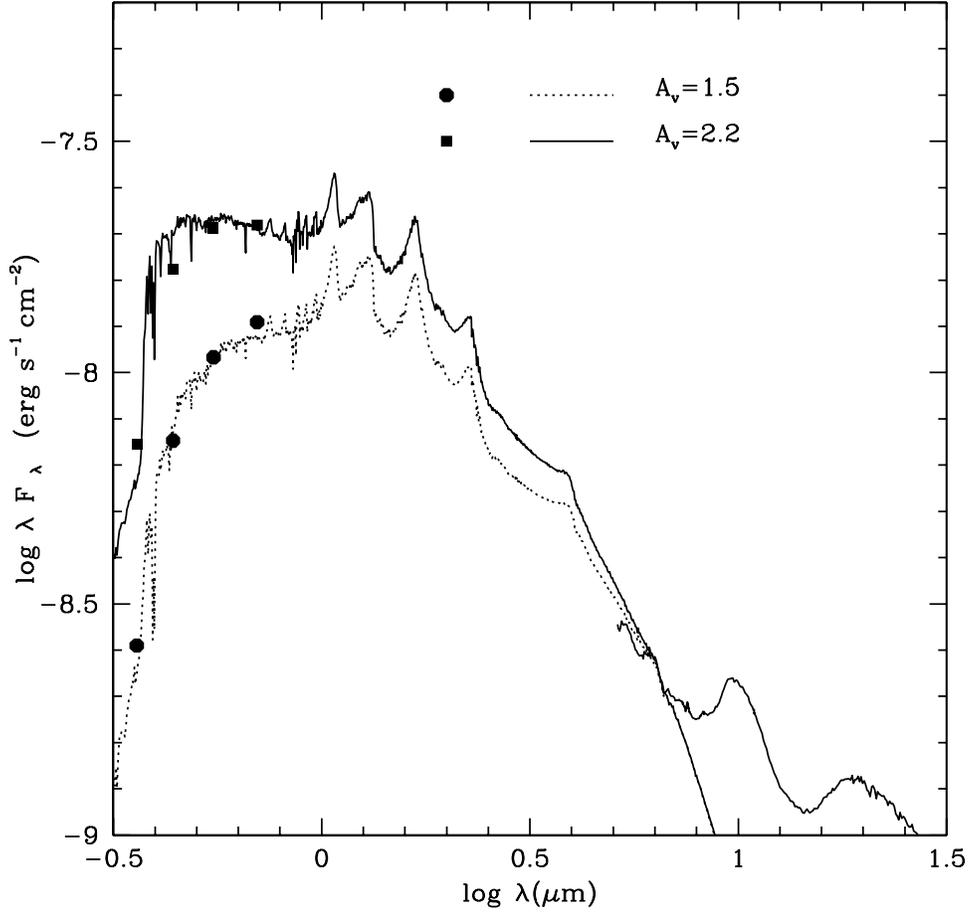} \caption{Two truncated models
fitting the observations corrected by A$_{V}$=2.2 and A$_{V}$=1.5.
Only optical photometries and IRS spectra (5-30 microns) are shown.
The square points are the photometries corrected by A$_{V}$=2.2,
while the round points are the photometries corrected by
A$_{V}$=1.5. The IRS spectra corrected by these two A$_{V}$ are
coincident. The solid line is the model fitting to observations with
A$_{V}$=2.2, while the dotted line is the model fitting to
observations with A$_{V}$=1.5. Both of the two models have the same
outer radius of the high state R$_{out}$=210 R$_{\odot}$.}
\label{fig:7}
\end{figure}

We explored the effects of a larger visual extinction
by scaling the observations to A$_{V}$=2.2.  As shown in
Figure \ref{fig:7}, this makes only a very slight difference to the
long-wavelength spectrum and does not change our estimate of the outer
radius.

\begin{figure}
\epsscale{.80}
\plotone{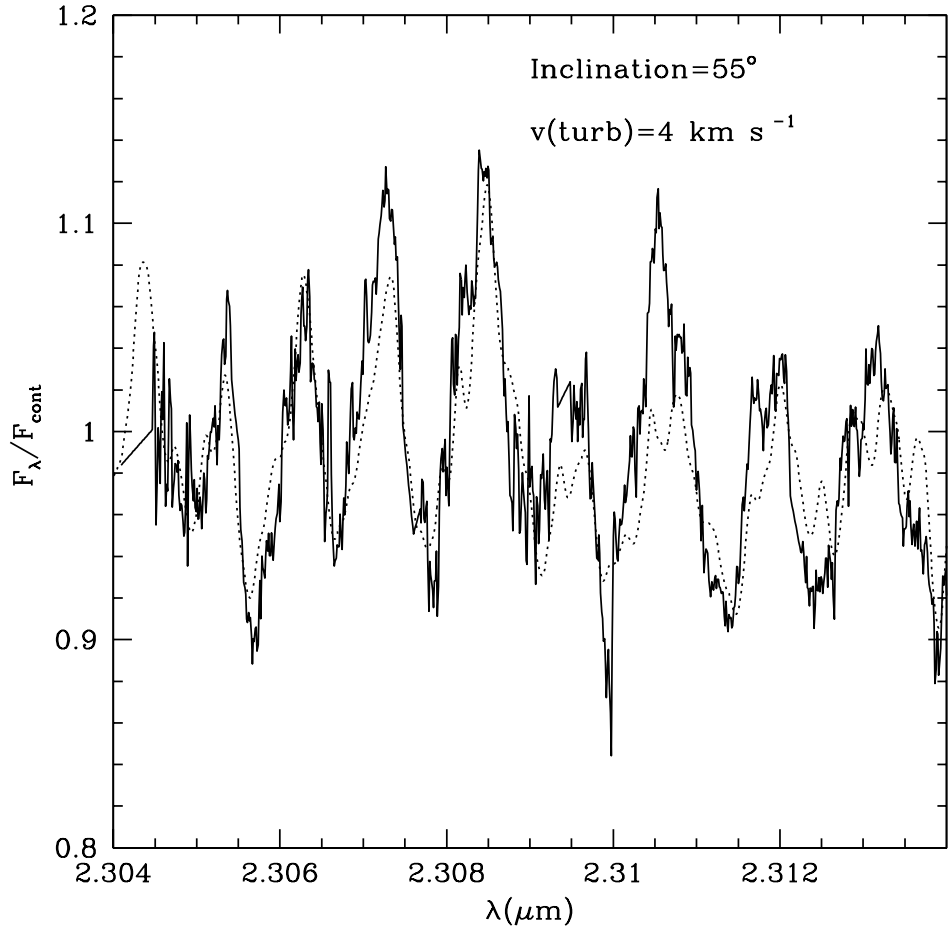}
\caption{CO first overtone absorption band around 2.3 $\mu$m.
The solid line is the observed CO first overtone absorption
feature from \cite{lee04} and the dotted line is the model spectrum after being
convolved to the same spectrum resolution. The turbulent velocity is 4 ${\rm km \, s^{-1}}$ and the inclination angle is 55 degrees.} \label{fig:CO}
\end{figure}

To examine our consistency with previous results,
we calculated the CO first overtone high resolution spectrum
with $R_{out}=210 R_{\odot}$ using SYNTHE program and compared it with the
high resolution spectrum from \cite{lee04}. The comparison is shown in Figure \ref{fig:CO}.
We found that we needed to adopt a turbulent velocity of 4 ${\rm km \, s^{-1}}$ to
obtain lines that are deep enough in comparison with the observations.
This is the sound speed of gas with temperature around 4000 K, slightly
supersonic for the annulus with radius larger than 4 stellar radius;
Hartmann, Hinkle, \& Calvet (2004) similarly found that slightly supersonic
turbulence was needed to explain the first-overtone CO lines, and pointed out
that this would not be surprising in the context of turbulence driven
by the magnetorotational instability.

\begin{figure}
\epsscale{.80} \plotone{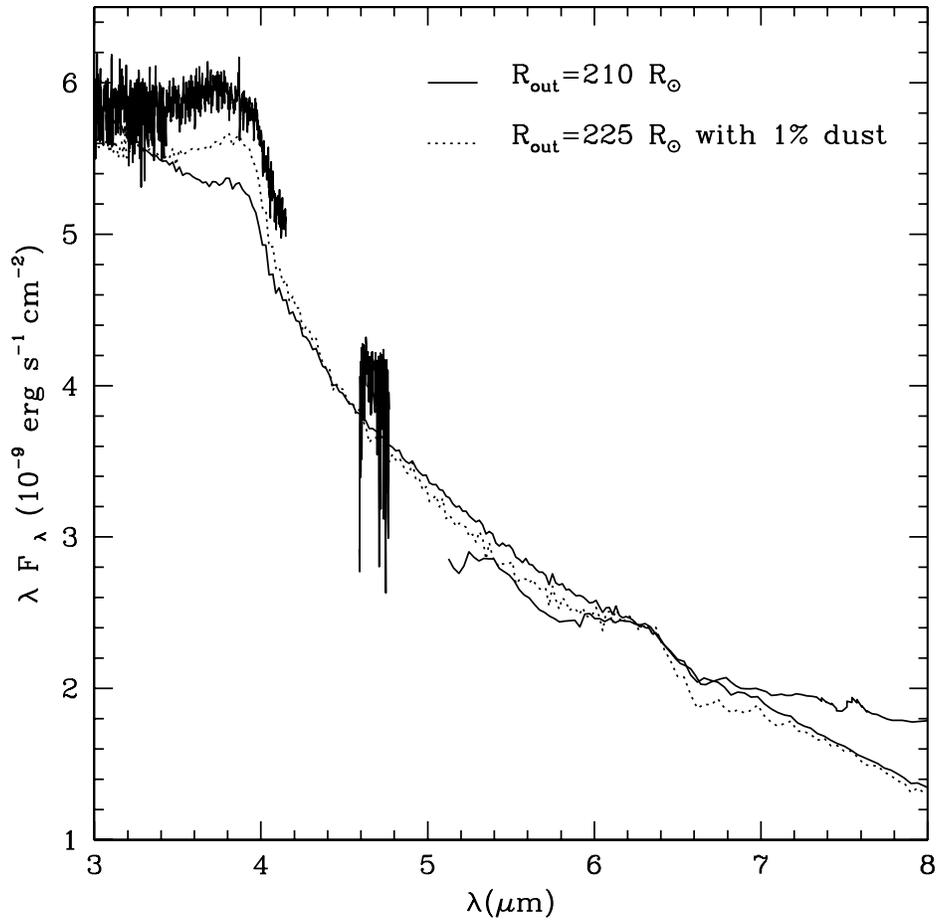} \caption{The mid-infrared part of
the observed spectra in Figure \ref{fig:1}  compared with the model
spectra. The solid line is the model with R$_{out}$=210 R$_{\odot}$
and the dotted line is the model with 1 \% dust opacity (dust
opacity
 reduced by a factor of 100) and R$_{out}$=225
R$_{\odot}$.
We do not have the CO fundamental opacities in our model and thus
cannot reproduce these strong spectral features at 4.6 $\mu$m.}
\label{fig:f9}
\end{figure}

In Figure \ref{fig:f9} we show an expanded view of the wavelength
region between $\sim 3$ and $8 \mu$m.  The model accounts for the
overall shape of the SED reasonably well, and yields a reasonable
strength for the 6.8$\mu$m water vapor absorption feature.  However,
the 5.8$\mu$m water vapor feature of the model is not as strong as
it is in the observations (\citep{jgreen06}).  In this wavelength
region the contributions from the dust-dominated regions of the disk
are significant, and the contribution of this dust (relatively
featureless) continuum emission could reduce the strength of the
absorption feature.  To investigate this possibility, we reduced the
dust opacity to only 1$\%$ that of
\cite{draine84,draine87} and recalculated the model (with a slightly
larger outer radius). As can be seen in Figure \ref{fig:f9}, this
reduction in dust opacity strengthens the absorption features,
although the predicted 5.8$\mu$m feature is still not strong enough,
perhaps suggesting some difficulty with the opacities.  It is not
clear whether we require dust depletion, because we have not
considered the temperature dependence of dust condensation in our
model.  We also cannot rule out uncertainties due to our
simple treatment of the vertical temperature structure of the disk.

\section{Discussion}

\subsection{The steady accretion disk model and the inner boundary condition}

We have shown that the steady accretion disk model can
quantitatively account for the observed SED of FU Ori over a factor
of nearly 20 in wavelength, from 4000 \AA\ to about $8 \mu$m.  This
is accomplished with only three adjustable parameters: $R_i$,
$\dot{M} M$, and $R_{out}$.  No alternative model proposed for FU
Ori has explained the SED over such a large wavelength range. The
accretion disk explanation is bolstered by this work.

 Within the
context of the steady disk model, however, there remains the
question of the appropriate inner boundary condition.  The steady
disk model considered here only accounts for half of the accretion
luminosity ($L = G \dot{M} M/(2 R_i)$) potentially available from
accretion onto a slowly rotating star.  As \cite{kenyon89} showed
from ultraviolet spectra, there is no evidence in FU Ori for
boundary layer emission which would account for this missing
luminosity, up to $\sim$ 10-20\% of the total luminosity; and if our
lower estimate of extinction is correct, the potential boundary
layer emission would be much less. This leaves as possibilities that
the missing accretion energy is going into spinning up the central
star, expanding the central star,  being radiated over a larger
radial distance of the disk, or some combination of all three
effects.

Hartmann \& Kenyon (1985; also Kenyon \etal 1989)
pointed out that the 5 R$_{\odot}$ inner radius implied by steady disk models is
considerably larger than that of the typical T Tauri star ($R \sim 2 \rsun$), and suggested that this might be due to the
disk dumping large amounts of thermal energy into the star, expanding its
outer layers.

Popham \etal (1993, 1996) suggested that the the missing boundary
layer energy could be distributed over a significant range of radii
in the disk at high accretion rates.  They found that as $\dot{M}$
increases to $\sim$ 10$^{-4}$ M$_{\odot}$yr$^{-1}$ the dynamical
boundary layer (the radial extent of the region where $\Omega$ drops
from Keplerian angular velocity to 0) grows to 10-20 $\%$ of a
stellar radius and the thermal boundary layer (the radial width over
which the boundary layer luminosity is radiated) grows to the point
that it becomes impossible to distinguish the boundary layer from
the inner parts of the disk. Popham \etal (1996) showed that this
model would imply somewhat slower than Keplerian rotation in the
innermost disk. In
this paper we find no evidence for extra radiation if A$_{V}$=1.5.
We cannot rule out a slightly larger A$_{V}$, which would allow some
extra dissipation of kinetic energy in the inner disk. However, the
shape of the spectrum in the blue-optical region (Figure \ref{fig:f5}) suggests
that the higher extinction value is not preferred, limiting the
amount of excess radiation to a modest fraction of the total
luminosity. We speculate that the missing energy is mostly going
into expanding the outer stellar layers, with perhaps a small amount
of excess radiation that is difficult to discern given the
uncertainties in the extinction.

\subsection{$R_{out}$ and $\alpha$}

Our estimate of the outer radius of the high state of FU Ori is around 200 R$_{\odot}$, although the
irradiation effect from the hot inner disk to the outer disk may decrease this value to no less than 100 $\rsun$.
\cite{bell94} assume low viscosity parameters ($\alpha$$\sim$10$^{-3}$)
and derive a small outer radius ( R$_{out}$$\sim$20R$_{\odot}$) for the high $\dot{M}$,
hot inner disk during outbursts.  Our model of FU Ori is inconsistent with such a small hot region; instead, we require the high $\dot{M}$ region to be an order of magnitude larger in radius.

FU Ori has been fading slowly in brightness over the last $\sim$~70 years \citep{ibrahimov99}.
If we attribute this fading to the emptying out of mass from the inner hot disk
onto the central star, this timescale gives us a way to constrain the viscosity in
this region.  We make the simple assumption that the decay timescale is the viscous timescale
\begin{equation}
t_v \sim R^{2}/\nu\,,
\end{equation}
where the viscosity is
$\nu = \alpha c_{s}^{2}/\Omega$, $\alpha$ is the viscosity parameter (Shakura \&
Sunyaev 1973), $c_s$ is the sound speed in the disk midplane, and $\Omega$ is the
Keplerian angular frequency at $R$.  We use the isothermal sound speed with a mean molecular
weight of 2.3, appropriate for molecular hydrogen.  We then need an estimate of the central
temperature to evaluate $t_v$.  Our radiative transfer
model shows that the effective temperature is $T_{eff} \sim 800$~K at the outer radius
of $R \sim 200 \rsun$. This is a lower limit for the central temperature as the absorption features in
the spectrum show that the central temperature is higher than the surface temperature.
The disk is likely to be very optically thick in this region,
such that the central temperature is considerably larger than the effective temperature
due to radiative trapping. We choose 1500 K as the central temperature
above which MRI is active and thus high $\alpha$ can sustain high mass accretion rate. We therefore find
\begin{equation}
t_v \sim 141\, {\rm yrs} \times \left (\frac{M}{0.3M_{\odot}}
\right)^{1/2} \left (\frac{R}{210R_{\odot}} \right )^{1/2} \left
(\frac{T}{1500 K} \right )^{-1} \left (\frac{\alpha}{10^{-1}}
\right )^{-1}\,. \label{eq:tvisc}
\end{equation}
where $M$ is the mass of the central star, $R$ is the outer radius
of the high mass accretion disk, $T$ is the central disk temperature
of the high mass accretion disk, and $\alpha$ is the viscosity
parameter.
If we estimate a typical decay timescale of FU Ori as $\sim 100$~yr, then we
find $\alpha \sim 0.14$. For the smaller hot disk radius with an
irradiated outer disk, our preliminary results yield (Zhu et al., in preparation) 
R $\ge$0.5 AU and thus $\alpha \sim 0.1$.

It is important to recognize that equation (\ref{eq:tvisc}) is only an order of magnitude
estimate.  The decay timescale of accretion can be affected by whatever mechanism makes
the transition from high to low state; for example, in the thermal instability model,
the decay time may not be the timescale for emptying out half the material in the
affected region of the disk, but only that (smaller) amount necessary to shut off the
thermal instability.  Nevertheless, it is suggestive that we find a larger value of
$\alpha$ in the high state than the $10^{-3}$ of \cite{bell94}.

A value
of $\alpha \sim 0.2 - 0.02$ is roughly consistent with estimates from
compact systems with accretion disks \citep{king2007} as well as
with numerical simulations of the magnetorotational instability (MRI) in ionized
disks (Balbus \& Hawley 1988).  In this connection, we note that the
requirement for enough thermal ionization to initiate a robust MRI is
a central temperature $T \gtrsim 1000$~K (Gammie 1996), comparable to
our observationally-based estimates ($T_{eff}$$\sim$800 K at $R$ $\sim$ 200 $\rsun$).

\subsection{Thermal and other instabilities}

The thermal instability model was originally suggested to account for outbursts
in dwarf nova systems (Faulkner, Lin, \& Papaloizou 1983) and has attractive
properties for explaining FU Ori outbursts \citep{bell94}.
However, the observed large outer radius, 0.5 $-$ 1 AU,
of the high accretion rate portion of the FU Ori disk
poses difficulties for the pure thermal instability model.
Even in the situation explored by \cite{2004MNRAS.353..841L},
in which the thermal instability is triggered
by a massive planet in the inner disk, the outer radius
of the high state is still the same as in \cite{bell94}, $\sim$ 20 $\rsun$.
The relatively high temperatures ($2-4 \times 10^3$~K) required to trigger
the thermal instability (due to the ionization of hydrogen)
are difficult to achieve at large radii; they require large surface densities which in turn
produce large optical depths, trapping the radiation internally and making
the central temperature much higher than the surface (effective) temperature.

To illustrate the problem, we calculated thermal equilibrium ``S''
curves (e.g., Faulkner \etal 1983) for our disk parameters at $R =
210 \rsun$ but various values of $\alpha$.  The equilibrium curve
represents energy balance between viscous energy generation and
radiative losses,
\begin{equation} F_{vis}=F_{rad}
\end{equation}
Thus,
\begin{equation}
\frac{9}{4}\alpha\Omega\Sigma c_{s}^{2}=2\sigma T_{eff}^{4}\,,
\end{equation}
where $\Sigma$ is the surface density and $c_{s}$ is the sound speed
at the local central disk. For an optically-thick disk, we have
T$_{c}^{4}$=$\frac{3}{8}$$\tau_{R}$T$_{eff}^{4}$, where
$\tau_{R}$=$\kappa_{R}$$\Sigma$ and $\kappa$ is the Rosseland mean
opacity \citep{hubeny90}.
Then,
\begin{equation}
T_{c}^{3}=\frac{27}{64\mu\sigma}\alpha\Omega\Sigma^{2} {\cal{R}}_{c} \kappa
 \,,
\end{equation}
where $\cal{R}$$_{c}$ is the gas constant, and $\mu$ is the mean
molecular weight . As long as we know T$_{c}$, we can derive
T$_{eff}$ and the corresponding $\dot{M}$ at this radius. Because
the Rosseland mean opacity is also dependent on temperature and
pressure, iteration of the disk structure calculation is needed to
derive the equilibrium curve.

\begin{figure}
\epsscale{1.00} \plotone{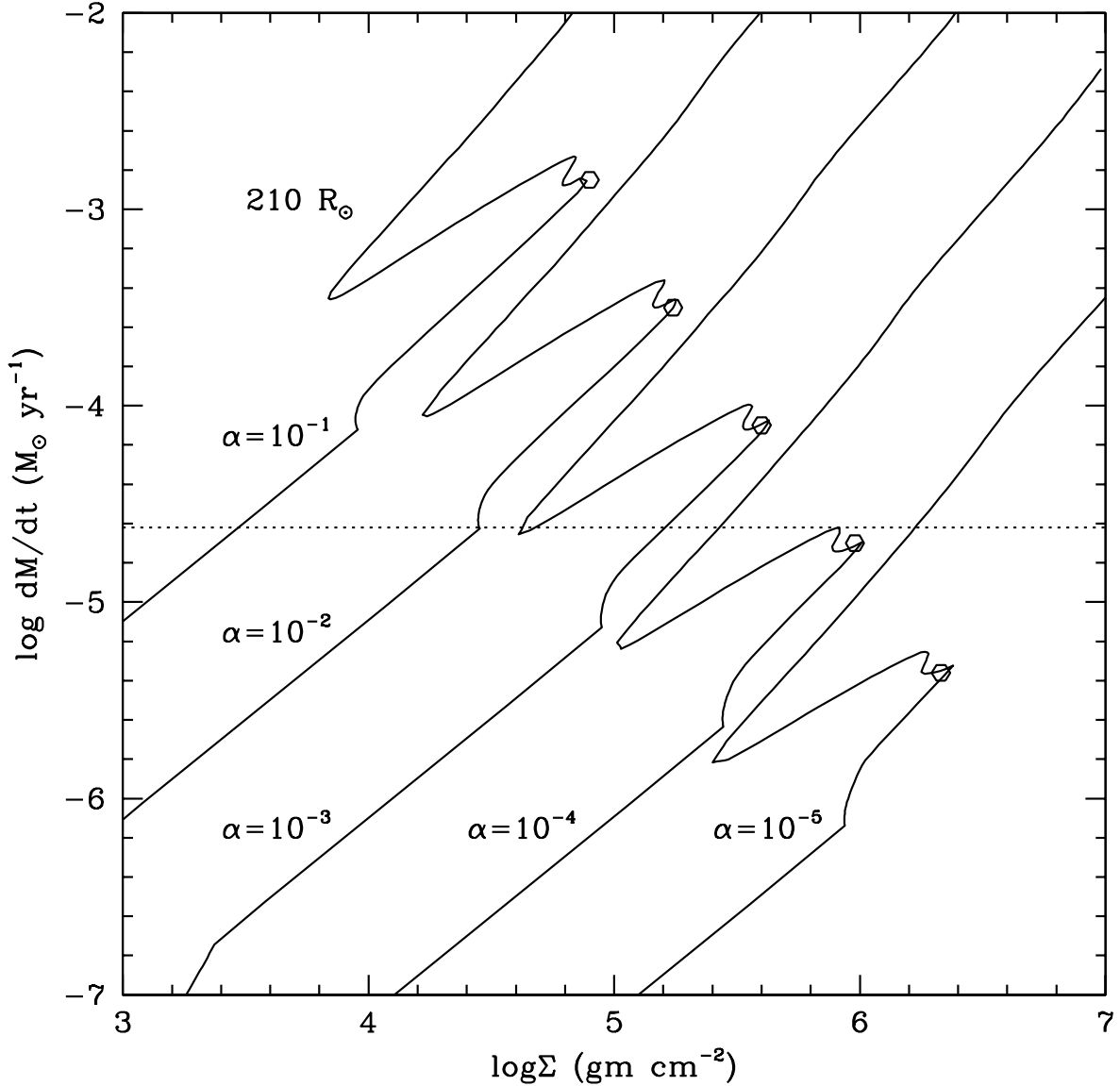} \caption{ The equilibrium curves
for thermal-instability at R=210 R$_{\odot}$ with five values of
$\alpha$: 10$^{-1}$,10$^{-2}$,10$^{-3}$,10$^{-4}$,10$^{-5}$(from top
to bottom).
The dotted line is $\dot{M}=2.4\times10^{-5}M_{\odot}/yr$, set to be
one order of magnitude lower accretion rate than the accretion rate
of the inner disk. If the outer disk $\dot{M}$ were one order of
magnitude higher than this (so that the outer disk and inner disk
accretion rates are the same) the mid-infrared flux would
be much larger than observed (see discussion in \S 4). The open
circles denote the critical surface densities $\Sigma_{A}$ for
triggering the thermal instability at each value of $\alpha$ (see \S
5.3) } \label{fig:f10}
\end{figure}

We show in Figure \ref{fig:f10} the equilibrium curve at R=210 R$_{\odot}$ for
five values of $\alpha$: 10$^{-1}$, 10$^{-2}$, 10$^{-3}$, 10$^{-4}$, 10$^{-5}$.
In discussing our results, we define $\Sigma_{A}$ to be the
highest stable low state for a given $\alpha$.
If at any radius the surface density increases above
$\Sigma_{A}$, the disk can no longer stay stable on the
lower branch and strong local heating begins.
If the $\alpha$ of the low state is the same as the $\alpha$ of the high state,
then $\alpha$$\ge$10$^{-2}$; in this case, to trigger the thermal instability at
$\Sigma_{A}$, the mass accretion rate in the low state would need to be $\sim$10$^{-3}$
(see Figure \ref{fig:f10}), which is higher than the $\dot{M}$ of the high state
 that we have determined. To have a mass accretion in the low state at least one order
of magnitude lower than that of the high state, we would need $\alpha$$\le$10$^{-4}$ for
the low state (see Figure \ref{fig:f10}).
This is comparable to the low state $\alpha$ used by \cite{bell94},
but very much lower than our high state estimate. Changing the outer radius to R$\sim$100 $\rsun$
does not change these results by more than a factor of 10.

With $\alpha$=10$^{-4}$, the critical surface density at 210 R$_{\odot}$
is $\Sigma_A \sim 10^{6}$ g cm$^{-2}$.  Then we can estimate the Toomre
$Q$ parameter,
\begin{equation}
Q = {c_s \Omega \over \pi G \Sigma_A} \sim 0.1 \,,
\end{equation}
where $Q < 1$ implies gravitational instability. Thus, a thermal instability model
for FU Ori implies a massive, probably gravitationally-unstable disk, which
raises the question: could the outburst be driven by gravity instead?

\cite{armitage01} proposed an alternative model of disk outbursts
in which material piles up at radii of order 1-2 AU, consistent with layered
disk models \citep{gammie96}, achieving repeated outbursts with high accretion
on size scales much closer to our inferred $R_{out}$.  In this model, gravitational
torques lead to increased accretion, which in turn heats up the inner disk
until temperatures are high enough (800 K in their model) to trigger the MRI,
which results in very much higher viscosities and rapid accretion.  The \cite{armitage01}
model does not give exactly FU Ori-type outbursts; instead, the accretion events
last for $\sim 10^{4}$ yr and achieve peak accretion rates of only $\sim 10^{-5} \msunyr$.
However, \cite{armitage01} suggested that thermal instabilities might be triggered
by their gravitationally-driven outbursts.  Simple models with somewhat different
parameters demonstrate that it is indeed possible to trigger thermal instabilities
in the context of this general model (Gammie, Book, \& Hartmann 2007, in preparation),
resulting in outbursts more similar to that of FU Ori.

Vorobyov \& Basu (2005,2006) suggest FU Ori outburst could be
explained by the accretion of clumps formed in a gravitationally
unstable disk. The disk becomes gravitationally unstable because of
the infall of the matter from the envelope. The spiral arms and
clumps form and grow in the unstable disk. The redistribution of
mass and angular momentum leads to centrifugal disbalance in the
disk and further triggers the outburst when the dense clumps are
driven into the central star. The authors derive a mass accretion
rate at 10 AU of $\sim$10$^{-4}$M$_{\odot}$/yr, which could result in
too much emission from the outer disk; this issue requires further
exploration.

The rapid rise time of the outburst of FU Ori in the B band, $\sim$ 1 year,
provides the best evidence for a thermal instability.
We can test this idea very crudely by using our steady disk model to
estimate the range of radii in the disk which can contribute to the B band
magnitude in the light curve, finding that only the inner
$R \lesssim 20 \rsun$ disk are involved.
If we then assumed that the rise in B light is due to the inward
propagation of a thermal instability front across this region at a
(maximum) speed $\sim \alpha c_s$ \citep{lin85,bell95}, and
assume a sound speed characteristic of $2-4 \times 10^3$~K
(the minimum temperature for instability), then we
derive $\alpha \sim 0.1$.
While this calculation is very crude, it does suggest consistency with the
idea of a thermal instability with
a viscosity parameter comparable to that which we inferred from the
decaying light curve (\S 5.2).

Finally, we estimate the amount of mass in the inner disk.
An upper mass limit for the inner disk can be estimated by assuming Q
 $\sim 1$ (the approximate limit for strong gravitational instability)
 at $R$ $\sim$ 200 $\rsun$. Adopting a central temperature $T \sim 1000 $~K,
the gravitational instability limit leads to an approximate surface density
limit $\Sigma \lesssim 10^{5} g cm^{-2}$ and thus a maximum inner disk mass
$\sim \pi\Sigma R^{2} \sim 0.05 \msun$. This leads us to an inner disk mass
estimate $0.05 \msun \gtrsim M_{in} \gtrsim 0.01 \msun$, where the lower limit
comes from requiring the outburst mass accretion rate of $\sim 10^{-4} \msunyr$
to be sustained for the observed decay timescale of $\sim 100$~yr.

\section{Conclusions}

Using the latest opacities and the ODF method, we have constructed a
new detailed radiative transfer disk model which reproduces the main
features of the FU Ori spectrum from $\sim$4000$\dot{A}$ to about 8
$\mu$m. The SED of FU Ori can be fitted with a steady disk model and
no boundary layer emission if A$_{V}$$\sim$1.5, a somewhat
lower valued extinction than
previously estimated. A larger A$_{V}$ would imply extra heating at
small radii which would result from dissipation of kinetic energy as
material accretes onto the central star. With A$_{V}$=1.5, the
inclination estimated from near-IR interferometry and the
observed rotation of the inner disk, we estimate that the central
star has a mass $\sim 0.3 M_{\odot}$ (typical of low-mass TTS) and a disk mass
accretion rate of $\sim 2.4 \times 10 ^{-4}\msunyr$.

Relying on the Infrared Spectrograph Spectrum of FU Ori from
the Spitzer Space Telescope presented by \cite{jgreen06}, we estimate that
the outer radius of the hot, rapidly accreting region of the inner disk
is $\sim$1 AU, although the irradiation effect may decrease this value by
no more than a factor of two(Zhu et al., in preparation).
Either way, this is inconsistent with the pure thermal instability
models of \cite{bell94} who adopted a low $\alpha$ viscosity parameter
$\sim$10$^{-3}$-10$^{-4}$.
If we assume that the observed decay timescale of
FU Ori ($\sim$100 years) is the viscous transport time from the outer edge of the hot region,
then we derive $\alpha \sim 0.2 - 0.02$, , comparable to the predictions of simulations
of the magnetohydrodynamic (MHD) turbulence. The effective temperature
at the outer edge of the hot region in our model is $\sim$800 K
; the presence of absorption features shows that the central disk temperature must be
higher. Thus, our model suggests that the central disk temperature is greater than
the T$_{c}$$\sim$1000 K required to maintain full MHD turbulence by thermal
ionization out to $\sim$ 1 AU.

We show that pure thermal instability models
have difficulty in explaining the outburst of FU Ori, and suggest that models
including gravitational torques and MRI activation,
such as those of \cite{armitage01}, are more promising.

\acknowledgments We thank Dr. R. Kurucz and Dr. F. Castelli for making
their line list and ATLAS codes available. They also provide us very
helpful suggestions to help us run their codes smoothly. We also
thank Mansur Ibrahimov for communicating recent photometry of FU Ori
and Thomas P. Greene for providing the near-infrared spectrum of FU
Ori.  This work was supported in part by NASA grant NNG06GJ32G and the
University of Michigan.

\clearpage
\begin{table}
\begin{center}
\caption{Best fit model \label{tab1}}
\begin{tabular}{clllll}

\tableline\tableline
M$_{star}$/M$_{\odot}$  & $\dot{M}$ & R$_{i}$/R$_{\odot}$ & R$_{out}$/R$_{\odot}$ & $\theta$\tablenotemark{a} & Distance/pc \\
\tableline
0.3 &2.4$\times$10$^{-4}$M$_{\odot}$/yr &5 &210 &55$^{o}$ &500 \\
\tableline
\end{tabular}
\tablenotetext{a}{Inclination angle of the disk from \cite{malbet05}}
\end{center}
\end{table}
\clearpage

\begin{table}
\begin{center}
\caption{Optical lines used to determine the spectral type \label{tab2}}
\begin{tabular}{clcl}
\tableline\tableline
Lines \tablenotemark{a}& Metals & Spectral type \\
\tableline
4047.00 & Fe I+Sc I & F8 \\
4226.00 & Ca I & G1 \\
4271.00 & Fe I & G4 \\
4305.00 & CH(Gband) & G1 \\
4458.00 & Mn I+Fe I & G1 \\
5329.00 & Fe I & G2 \\
5404.00 & Fe I & G3 \\
6162.00 & Ca I+TiO & F9 \\
\tableline
\end{tabular}
\tablenotetext{a}{From \cite{jesusphd}}
\end{center}
\end{table}

\clearpage

\clearpage

\begin{thebibliography}

\bibitem[Alexander \& Ferguson(1994)]{alexander94} Alexander,
D.~R., \& Ferguson, J.~W.\ 1994, \apj, 437, 879

\bibitem[Armitage et al.(2001)]{armitage01} Armitage, P.~J., Livio, M., \& Pringle, J.~E.\ 2001, \mnras, 324, 705

\bibitem[Aspin \& Reipurth(2003)]{aspin03} Aspin, C., \&
Reipurth, B.\ 2003, \aj, 126, 2936

\bibitem[Balbus \& Hawley(1998)]{1998RvMP...70....1B} Balbus, S.~A., \&
Hawley, J.~F.\ 1998, Reviews of Modern Physics, 70, 1

\bibitem[Bell \& Lin(1994)]{bell94} Bell, K.~R., \& Lin,
D.~N.~C.\ 1994, \apj, 427, 987

\bibitem[Bell et al.(1995)]{bell95} Bell, K.~R., Lin,
D.~N.~C., Hartmann, L.~W., \& Kenyon, S.~J.\ 1995, \apj, 444, 376

\bibitem[Calvet et al.(1991b)]{calvet91b} Calvet, N., Hartmann,
L., \& Kenyon, S.~J.\ 1991, \apj, 383, 752


\bibitem[Calvet et al.(1991a)]{calvet91} Calvet, N., Patino, A.,
Magris, G.~C., \& D'Alessio, P.\ 1991, \apj, 380, 617

\bibitem[Castelli(2005)]{castelli05} Castelli, F.\ 2005, Memorie
della Societa Astronomica Italiana Supplement, 8, 34

\bibitem[Castelli \& Kurucz(2004)]{kurucz04} Castelli, F., \&
Kurucz, R.~L.\ 2004, ArXiv Astrophysics e-prints, arXiv:astro-ph/0405087

\bibitem[Clarke et al.(2005)]{clarke05} Clarke, C., Lodato, G.,
Melnikov, S.~Y., \& Ibrahimov, M.~A.\ 2005, \mnras, 361, 942

\bibitem[Cohen et al.(2003)]{cohen03} Cohen, M., Wheaton,
W.~A., \& Megeath, S.~T.\ 2003, \aj, 126, 1090

\bibitem[Cohen \& Woolf(1971)]{cohen71} Cohen, M., \& Woolf,
N.~J.\ 1971, \apj, 169, 543

\bibitem[Draine \& Lee(1984)]{draine84} Draine, B.~T., \& Lee,
H.~M.\ 1984, \apj, 285, 89

\bibitem[Draine \& Lee(1987)]{draine87} Draine, B.~T., \& Lee,
H.~M.\ 1987, \apj, 318, 485

\bibitem[Fabricant et al.(1998)]{fabricant98} Fabricant, D.,
Cheimets, P., Caldwell, N., \& Geary, J.\ 1998, \pasp, 110, 79

\bibitem[Faulkner et al.(1983)]{1983MNRAS.205..359F} Faulkner, J., Lin,
D.~N.~C., \& Papaloizou, J.\ 1983, \mnras, 205, 359

\bibitem[Ferguson et al.(2005)]{ferguson05} Ferguson, J.~W.,
Alexander, D.~R., Allard, F., Barman, T., Bodnarik, J.~G., Hauschildt,
P.~H., Heffner-Wong, A., \& Tamanai, A.\ 2005, \apj, 623, 585

\bibitem[Gammie(1996)]{gammie96} Gammie, C.~F.\ 1996, \apj, 457,
355

\bibitem[Gammie(1999)]{gammie99} Gammie, C.~F.\ 1999, ASP
Conf.~Ser.~160: Astrophysical Discs - an EC Summer School, 160, 122

\bibitem[Green et al.(2006)]{jgreen06} Green, J.~D., Hartmann, 
L., Calvet, N., Watson, D.~M., Ibrahimov, M., Furlan, E., Sargent, B., \& 
Forrest, W.~J.\ 2006, \apj, 648, 1099 

\bibitem[Greene \& Lada(1996)]{greene96} Greene, T.~P., \& Lada,
C.~J.\ 1996, \aj, 112, 2184

\bibitem[Goodrich(1987)]{goodrich87} Goodrich, R.~W.\ 1987, \pasp,
99, 116

\bibitem[Hartmann et al.(2004)]{lee04} Hartmann, L., Hinkle,
K., \& Calvet, N.\ 2004, \apj, 609, 906


\bibitem[h (1985)]{hk85} Hartmann, L., \& Kenyon, S.~J.\ 1985, \apj, 299, 462

\bibitem[h (1987)a]{hk87a} Hartmann, L., \& Kenyon, S.~J.\ 1987a, \apj, 312, 243

\bibitem[h (1987)b]{hk87b} Hartmann, L., \&
Kenyon, S.~J.\ 1987b, \apj, 312, 243

\bibitem[Hartmann \& Kenyon(1991)]{HK91} Hartmann, L., \&
Kenyon, S.~J.\ 1991, IAU Colloq.~129: The 6th Institute d'Astrophysique de
Paris (IAP) Meeting: Structure and Emission Properties of Accretion Disks,
203

\bibitem[Hartmann \& Kenyon(1996)]{Lee96} Hartmann, L., \&
Kenyon, S.~J.\ 1996, \araa, 34, 207

\bibitem[Herbig(1977)]{herbig77} Herbig, G.~H.\ 1977, \apj, 217,
693

\bibitem[Hernandez(2005)]{jesusphd} Hern\'{a}ndez, J.\ 2005, Ph.D. Thesis, Postgrado de F\'{\i}sica Fundamental, Universidad
de Los Andes, Venezuela.

\bibitem[Hern{\'a}ndez et al.(2004)]{jesus04} Hern{\'a}ndez,
J., Calvet, N., Brice{\~n}o, C., Hartmann, L., \& Berlind, P.\ 2004, \aj,
127, 1682

\bibitem[Hubeny(1990)]{hubeny90} Hubeny, I.\ 1990, \apj, 351,
632

\bibitem[Ibrahimov(1999)]{ibrahimov99} Ibrahimov, M.~A.\ 1999,
Informational Bulletin on Variable Stars, 4691, 1

\bibitem[Jacoby et al.(1984)]{jacoby84} Jacoby, G.~H., Hunter,
D.~A., \& Christian, C.~A.\ 1984, \apjs, 56, 257

\bibitem[Kenyon \& Hartmann(1991)]{kenyon91} Kenyon, S.~J., \&
Hartmann, L.~W.\ 1991, \apj, 383, 664

\bibitem[Kenyon et al.(1993)]{kenyon93} Kenyon, S.~J., Hartmann,
L., Gomez, M., Carr, J.~S., \& Tokunaga, A.\ 1993, \aj, 105, 1505

\bibitem[Kenyon et al.(1988)]{kenyon88} Kenyon, S.~J., Hartmann,
L., \& Hewett, R.\ 1988, \apj, 325, 231 (KHH88)

\bibitem[Kenyon et al.(1989)]{kenyon89} Kenyon, S.~J., Hartmann,
L., Imhoff, C.~L., \& Cassatella, A.\ 1989, \apj, 344, 925

\bibitem[King et al.(2007)]{king2007} King, A.~R., Pringle, 
J.~E., \& Livio, M.\ 2007, \mnras, 376, 1740

\bibitem[Kurucz(1993)]{kurucz93} Kurucz, R.~L.\ 1993, Kurucz
CD-ROM, Cambridge, MA: Smithsonian Astrophysical Observatory, |c1993,
December 4, 1993,

\bibitem[Kurucz(2005)]{kurucz05} Kurucz, R.~L.\ 2005, Memorie
della Societa Astronomica Italiana Supplement, 8, 86

\bibitem[Kurucz \& Avrett(1981)]{kurucz81} Kurucz, R.~L., \&
Avrett, E.~H.\ 1981, SAO Special Report, 391,

\bibitem[Kurucz et al.(1974)]{kurucz74} Kurucz, R.~L.,
Peytremann, E., \& Avrett, E.~H.\ 1974, Washington : Smithsonian
Institution : for sale by the Supt.~of Docs., U.S.~Govt.~Print.~Off.,
1974., 37

\bibitem[Lin et al.(1985)]{lin85} Lin, D.~N.~C., Faulkner,
J., \& Papaloizou, J.\ 1985, \mnras, 212, 105

\bibitem[Lodato \& Clarke(2004)]{2004MNRAS.353..841L} Lodato, G., \&
Clarke, C.~J.\ 2004, \mnras, 353, 841

\bibitem[Malbet et al.(2005)]{malbet05} Malbet, F., et al.\
2005, \aap, 437, 627

\bibitem[Mould et al.(1978)]{mould78} Mould, J.~R., Hall,
D.~N.~B., Ridgway, S.~T., Hintzen, P., \& Aaronson, M.\ 1978, \apjl, 222,
L123

\bibitem[Muzerolle et al.(2003)]{muzerolle03} Muzerolle, J.,
Calvet, N., Hartmann, L., \& D'Alessio, P.\ 2003, \apjl, 597, L149

\bibitem[Papaloizou et al.(1983)]{papaloizou83} Papaloizou, J.,
Faulkner, J., \& Lin, D.~N.~C.\ 1983, \mnras, 205, 487

\bibitem[Partridge \& Schwenke(1997)]{PS97} Partridge, H.,
\& Schwenke, D.~W.\ 1997, \jcp, 106, 4618

\bibitem[Popham et al.(1993)]{popham93} Popham, R., Narayan, R.,
Hartmann, L., \& Kenyon, S.\ 1993, \apjl, 415,

\bibitem[Popham et al.(1996)]{1996ApJ...473..422P} Popham, R., Kenyon, S.,
Hartmann, L., \& Narayan, R.\ 1996, \apj, 473, 422

\bibitem[Quanz et al.(2006)]{quanz2006} Quanz, S.~P., Henning,
T., Bouwman, J., Ratzka, T., \& Leinert, C.\ 2006, \apj, 648, 472L127

\bibitem[Rayner et al.(1998)]{rayner98} Rayner, J.~T., Toomey,
D.~W., Onaka, P.~M., Denault, A.~J., Stahlberger, W.~E., Watanabe, D.~Y.,
\& Wang, S.-I.\ 1998, \procspie, 3354, 468

\bibitem[Reipurth \& Aspin(1997)]{reipurth97} Reipurth, B., \&
Aspin, C.\ 1997, \aj, 114, 2700

\bibitem[Reipurth \& Aspin(2004)]{reipurth2004} Reipurth, B., \&
Aspin, C.\ 2004, \apjl, 608, L65

\bibitem[Sandell \& Aspin(1998)]{sandell98} Sandell, G., \&
Aspin, C.\ 1998, \aap, 333, 1016

\bibitem[Sandell \& Weintraub(2001)]{2001ApJS..134..115S} Sandell, G., \&
Weintraub, D.~A.\ 2001, \apjs, 134, 115

\bibitem[Sbordone et al.(2004)]{kurucz042} Sbordone, L.,
Bonifacio, P., Castelli, F., \& Kurucz, R.~L.\ 2004, Memorie della Societa
Astronomica Italiana Supplement, 5, 93

\bibitem[Schwenke(1998)]{S98} Schwenke, D.~W.\ 1998,
Chemistry and Physics of Molecules and Grains in Space.~ Faraday
Discussions No.~109, 321

\bibitem[Shakura \& Sunyaev(1973)]{SS1973} Shakura, N.~I., \&
Sunyaev, R.~A.\ 1973, \aap, 24, 337

\bibitem[Simon et al.(2000)]{2000ApJ...545.1034S} Simon, M., Dutrey, A., \&
Guilloteau, S.\ 2000, \apj, 545, 1034

\bibitem[Turner et al.(1997)]{turner97} Turner, N.~J.~J.,
Bodenheimer, P., \& Bell, K.~R.\ 1997, \apj, 480, 754

\bibitem[Vorobyov \& Basu(2005)]{2005ApJ...633L.137V} Vorobyov, E.~I., \&
Basu, S.\ 2005, \apjl, 633, L137

\bibitem[Vorobyov \& Basu(2006)]{2006ApJ...650..956V} Vorobyov, E.~I., \&
Basu, S.\ 2006, \apj, 650, 956


\end{thebibliography}
\end{document}